\documentclass[aps,prd,showpacs,floatfix,preprintnumbers,nofootinbib,
amssymb,amsmath]{revtex4}
\usepackage{subfigure}
\usepackage{graphicx,bm,color}
\usepackage{amssymb}

\newcommand{\beq}{\begin{equation}}
\newcommand{\eeq}{\end{equation}}
\newcommand{\bea}{\begin{eqnarray}}
\newcommand{\eea}{\end{eqnarray}}
\newcommand{\bali}{\begin{align}}
\newcommand{\eali}{\end{align}}

\begin{document}
 

\title{Study of Beta Equilibrated 2+1 Flavor Quark Matter in PNJL Model}

\author{Abhijit Bhattacharyya}
\email{abphy@caluniv.ac.in}
\affiliation{Department of Physics, University of Calcutta,
92, A.P.C Road, Kolkata-700009, India}

\author{Sanjay K Ghosh}
\email{sanjay@bosemain.boseinst.ac.in}

\author{Sarbani Majumder}
\email{sarbanimajumder@gmail.com}

\author{Rajarshi Ray}
\email{rajarshi@bosemain.boseinst.ac.in}
\affiliation{Center for Astroparticle Physics \&
Space Science, Block-EN, Sector-V, Salt Lake, Kolkata-700091, INDIA 
 \\ \& \\ 
Department of Physics, Bose Institute, \\
93/1, A. P. C Road, Kolkata - 700009, INDIA}


\begin{abstract}

We report a first case study of the phase diagram of 2+1 flavor strongly
interacting matter in $\beta-$equilibrium, using the
Polyakov$-$Nambu$-$Jona-Lasinio model. Physical characteristics of
relevant thermodynamic observables have been discussed. A comparative
analysis with the corresponding observables in the
Nambu-Jona-Lasinio model is presented. We find distinct differences
between the models in terms of a number of thermodynamic quantities
like the speed of sound, specific heat, various number densities as
well as entropy. The present study is expected to give us a better
insight into the role that the superdense matter created in heavy ion
collision experiments play in our understanding of the properties of
matter inside the core of supermassive stars in the Universe.

\end{abstract}
\maketitle

\section{Introduction} \label{sc.intro}

The phase diagram of strongly interacting matter has been at the center
of attention for quite some time now. Under a variety of extreme 
conditions of temperature and/or density the hadrons may overlap and
loose their individuality and a new state of matter called Quark Gluon
Plasma (QGP) may be formed \cite{muller}. It is well believed that such
a state of matter existed in the hot early Universe, a few microsecond
after the Big Bang. Deconfined quark matter could also exist in the core
of neutron stars (NS)
\cite{rajagopal-condmatt, blaschke-2010, glendenning3}
where the temperature is relatively low but density is high. So an
understanding of the physics of strongly interacting matter at such
environmental conditions would have important cosmological and
astrophysical significance.

In the laboratory such conditions of large temperatures and densities
can be created by the collision of heavy ions at high energies.
Presently the strongly interacting matter at high temperature and close
to zero baryon densities $-$ a scenario relevant for early universe $-$
is being explored at Relativistic Heavy Ion Collider (RHIC) at BNL and
the Large Hadron Collider (LHC) at CERN. A wealth of information has
been obtained from RHIC, and a lot more is expected from both the
future runs there as well as from LHC. More recently a variety of energy
scans at RHIC and the upcoming facility (FAIR) at GSI, are expected to
give us a glimpse of matter in the baryon-rich environments - the
so-called {\it compressed baryonic matter} (CBM). These experiments will
also be useful in the search for signatures of critical phenomena
associated with a second order critical end point (CEP).

At the same time, observational data are being collected by a large
number of telescopes and satellites \cite{bielich} such as the radio
telescopes at the Arecibo, Parkes, Jo-drell Bank, and Green Bank
Observatories, the Hubble Space Telescope, European Space Agency's
International Gamma Ray Astrophysics Laboratory (INTEGRAL) satellite,
Very Large Telescope (VLT) of the European Southern Observatory, the
X-ray satellites Chandra, XMM-Newton and NASA's Rossi X-ray Timing
Explorer and the Swift satellite. Observations from these facilities
are supposed to tell us about the properties of strongly interacting
matter at high densities relevant for the astrophysics of compact stars.

Thus on one hand the laboratory experiments are expected to scan the
phase space temperature and various conserved quantum number densities
of strongly interacting matter. On the other hand the astrophysical
observations are expected to uncover the physics for high baryon 
number density region of the phase diagram. It should be noted here
that the physical characteristics of the matter under consideration may
be quite different in the two cases. The time-scale of the dynamics of
heavy-ion experiments is so small that only strong interactions may
equilibrate thermodynamically. While the dynamics in the astrophysical
scenario is slow enough to allow even weak interactions may equilibrate.
Thus a question naturally arises $-$ to what extent can laboratory
experiments be used to infer about the compact star interiors? The aim
of this paper is to address this question at a preliminary level from
the characteristics of the "$\beta-$equilibrated" phase diagrams.

One should be able to study the properties of systems described above
from first principles using Quantum Chromodynamics (QCD), which is
{\it the} theory of strong interactions. However, QCD is highly
non-perturbative in the region of temperature and density that we are
interested in. The most reliable way to analyze the physics in this
region of interest is to perform a numerical computation of the lattice
version of QCD (Lattice QCD). The scheme is robust but numerically
costly. Moreover, there are problems in applying this scheme for the
systems having finite baryon density. Thus it has become a common
practice to study the physics of strongly interacting matter under
the given conditions using various QCD inspired effective models.

Until now various quark models, such as, different versions of the
MIT bag model \cite{burgio,alford1}, the color-dielectric model
\cite{ghosh_star1,ghosh_star2} and different formulations of the NJL
model \cite{buballa,baldo1} have been used to study the NS structure.
Despite the similarity of the results on the value of the maximum NS
mass, the predictions on the NS configurations can differ substantially
from model to model. The most striking difference is in the quark matter
content of the NS, which can be extremely large in the case of EOS
related to the MIT bag model or the color-dielectric model, but it is
vanishingly small in the case of the original version of the NJL model
\cite{buballa,schertler}. In the case of NJL model it turns out that,
as soon as quark matter appears at increasing NS mass, the star becomes
unstable, with only the possibility of a small central region with a
mixed phase of nucleonic and quark matter. This may be a result of the
lack of confinement in NJL model. In fact an indirect relationship
between confinement and NS stability has been found in a study using NJL
model with density dependent cut-off \cite{baldo2}. Hence it is
important to study the EOS from the Polyakov$-$Nambu$-$Jona-Lasinio
(PNJL) model \cite{Ratti1,ghosh-prd73,Ghosh:2008}, where a better
description of confinement has been incorporated through Polyakov loop
mechanism. Moreover, a comparison with NJL model might be helpful in
understanding the role of Polyakov loop at high chemical potential.

A detailed study of 2+1 flavor strong interactions have been done by
some of us using the PNJL model. The general thermodynamic properties
along with the phase diagram \cite{paramita}, as well as details of
fluctuation and correlations of various conserved charges \cite{anirban}
have been reported. Here we extend the work by including
$\beta-$equilibrium into the picture. In the context of NJL model such a
study was done earlier in \cite{buballa-prd,hanauske-njl}.
The properties of pseudoscalar
and neutral mesons have been studied
in finite density region within the framework of 2+1 flavor NJL model in
$\beta-$equilibrium\cite{costa_meson1,costa_meson2}.
 
We investigate and compare different properties of the NJL and PNJL models
in the T-$\mu_B$ plane. The specialization of these studies to the
possible dynamical evolution of NS and/or CBM created in heavy-ion
collisions will be kept as a future excersize.

The paper is organized as follows. In section \ref{sc.formalism} we
discuss our model. In section \ref{sc.results} we calculate different
thermodynamic properties and present our result and finally we conclude
in section \ref{sc.conclusion}.

\section{Formalism}
\label{sc.formalism}

The supermassive compact objects like neutron stars are born in
the aftermath of supernova explosions. The initial 
temperature of a new born NS can be as high as $T \sim 100$ MeV. For
about one minute following its birth, the star stays in a special
proto-neutron star state: hot, opaque to neutrinos, and larger than an
ordinary NS (see, e.g., \cite{pons,yakov} and references therein). Later
the star becomes transparent to neutrinos generated in its interior. It
cools down gradually, initially through neutrino emission
($t \leq 10^5$ years) and then through the emission of photons
($t \geq 10^5$ years) \cite{yakovlev}, and transforms into an ordinary
NS. The weak interaction responsible for the emission of these neutrinos
eventually drive the stars to the state of $\beta-$equilibrium along
with the imposed condition of charge neutrality.

The mass, radius and other characteristics of such a star depend on the
equation of state (EOS), which in turn, is determined by the composition
of the star \cite{lattimer1}. The possible central density of a compact
star may be high enough for the usual neutron-proton matter to undergo a
phase transition to some exotic forms of strongly interacting matter.
Some of the suggested exotic forms of strongly interacting matter are
the hyperonic matter, the quark matter, the superconducting quark matter
etc. If there is a hadron to quark phase transition inside the NS, then
all the characteristics of the NS will depend on the nature of the phase
transition \cite{mishustin1,mishustin2}.

Furthermore, there have been suggestions that the strange quark matter,
containing almost equal numbers of u, d and s quarks, may be the ground
state of strongly interacting matter (see \cite{witten} and references
therein). If such a conjecture
is true, then there is a possibility of the existence of self-bound pure
strange stars as well. In fact, the conversion of NS to strange star may
really be a two step process \cite{mallik1}. The first process involves
the deconfinement of nuclear to two-flavor quark matter; the second
process deals with the conversion of excess down quarks to strange
quarks resulting into a $\beta-$equilibrated charge neutral strange
quark matter. There are several mechanisms by which the conversion of
strange quark may be triggered at the center of the star
\cite{alcock, glendenning-prd46}. The dominant reaction mechanism by
which the strange quark production in quark matter occurs is the
non-leptonic weak interaction process \cite{ghosh-npa}

\begin{equation}
 u_1+d\leftrightarrow u_2+s
\end{equation}

\noindent
Initially when the quark matter is formed, $\mu_d > \mu_s$, and the
above reaction converts excess d quarks to s quarks. But in order to
produce chemical equilibrium the semileptonic interactions,

\begin{equation}
d(s)\rightarrow u+e^{-}+\bar{{\nu}_e}
\end{equation}
\begin{equation}
 u+e^{-}\rightarrow d(s)+{\nu}_e
\end{equation}

\noindent
play important role along with the above non-leptonic interactions.
These imply the $\beta-$equilibrium condition
$\mu_d$=$\mu_u$+$\mu_e$+$\mu_{\bar{\nu}}$; and $\mu_s$=$\mu_d$.

Actually, the only conserved charges in the system are the baryon number
$n_B$ and the electric charge $n_Q$. Since we are assuming neutrinos to
leave the system, lepton number is not conserved \cite{buballa}. Strange
chemical potential $\mu_S$ is zero because strangeness is not conserved.
So two of the four chemical potentials ($\mu_u$, $\mu_d$, $\mu_s$ and
$\mu_e$) are independent. In terms of the baryon chemical potential
($\mu_B$), which is equivalent to the quark chemical potential
($\mu_q$=$\mu_B$/3), and the charge chemical potential ($\mu_Q$) these
can be expressed as,
$\mu_u=\mu_q+\frac{2}{3}\mu_Q$;  $\mu_d=\mu_q-\frac{1}{3}\mu_Q$;
$\mu_s=\mu_q-\frac{1}{3}\mu_Q$;  $\mu_e=-\mu_Q$.
These conditions are put as constraints in the description of the
thermodynamics of a given system through the PNJL model.

The thermodynamic potential of 2+1 flavor PNJL model for non-zero quark
chemical potential is \cite{paramita}

\begin{align}
\Omega &= {\cal {U^\prime}}[\Phi,\bar \Phi,T] +
 2{g_S}{\sum_{f=u,d,s}}{\sigma_f^2} -
 {\frac{g_D}{2}}{\sigma_u}{\sigma_d}{\sigma_s} -
 6{\sum_{f=u,d,s}}{\int_{0}^{\Lambda}}{\frac{d^3p}{(2\pi)^3}}
 E_{f}\Theta {(\Lambda-{ |\vec p|})}
\nonumber \\
& - 2T{\sum_{f=u,d,s}}{\int_0^\infty}{\frac{d^3p}{(2\pi)^3}}
 \ln\left[1+3(\Phi+{\bar \Phi}e^{-\frac{(E_{f}-\mu_f)}{T}})
    e^{-\frac{(E_{f}-\mu_f)}{T}} + e^{-3\frac{(E_{f}-\mu_f)}{T}}\right]
\nonumber \\
& - 2T{\sum_{f=u,d,s}}{\int_0^\infty}{\frac{d^3p}{(2\pi)^3}}
 \ln\left[1+3({\bar \Phi}+{ \Phi}e^{-\frac{(E_{f}+\mu_f)}{T}})
    e^{-\frac{(E_{f}+\mu_f)}{T}}+e^{-3\frac{(E_{f}+\mu_f)}{T}}\right]
\end{align}

\noindent
where, $\sigma_f=\langle{\bar \psi_f} \psi_f\rangle$ and
$E_{f}=\sqrt {p^2+M^2_f}$ with,
$M_f=m_f-2g_S\sigma_f+{\frac{g_D}{2}}\sigma_{f+1}\sigma_{f+2}$.

The effective potential $\mathcal{U}^{\prime}(\Phi,\bar{\Phi},T)$ is
expressed in terms of the traced Polyakov loop
$\Phi=(\mathrm{Tr}_c\,L)/N_c$ and its (charge) conjugate
$\bar{\Phi}=(\mathrm{Tr}_cL^{\dagger})/N_c$, where $L$ is a matrix in
color space given by, $L\left(\vec{x}\right)=\mathcal{P}
\exp\left[-i\int_{0}^{\beta}d\tau\,A_4\left(\vec{x},\tau\right)\right]$,
where $\beta = 1/T$ is the inverse temperature and
$A_4 = A_4^a \lambda_a$, $A_4^a$ being the temporal component of the 
Eucledian gluon field and $\lambda_a$ are the Gell-Mann matrices with
adjoint color indices $a=1,\cdots,8$. Assuming a constant $A_4^a$ and
the $A_i$'s to be zero for ($i=1,2,3$), $\Phi$ and its conjugate
$\bar{\Phi}$, are treated as classical field variables in PNJL model.
The temperature dependent effective potential
$\mathcal{U}^{\prime}(\Phi,\bar{\Phi},T)$ is so chosen to have exact
$Z(3)$ center symmetry and is given by,

\beq
\frac{\mathcal{U}^{\prime}(\Phi,\bar{\Phi})}{T^4} = 
\frac{\mathcal{U}(\Phi,\bar{\Phi})}{T^4} -
\kappa \ln [J(\Phi,\bar{\Phi})],
\label{eq.uup}
\eeq

\noindent
where,

\beq
\frac{\mathcal{U}\left(\Phi,\bar{\Phi},T\right)}{T^4} =
-\frac{b_2\left(T\right)}{2}\bar{\Phi} \Phi -
\frac{b_3}{6}\left(\Phi^3+ {\bar{\Phi}}^3\right) +
\frac{b_4}{4}\left(\bar{\Phi}\Phi\right)^2
\label{u1} 
\eeq 

\noindent
with $b_2\left(T\right) = a_0 + a_1\left(\frac{T_0}{T}\right) +
a_2\left(\frac{T_0}{T}\right)^2 + a_3\left(\frac{T_0}{T}\right)^3$,
and $J[\Phi,\bar{\Phi}] = (27/24\pi^2)
(1 - 6\,\bar{\Phi} \Phi + 4\,(\bar{\Phi}^3 + \Phi^3) -
3\,(\bar{\Phi} \Phi)^2)$ is the Vandermonde determinant. A fit of the
coefficients $a_i,~b_i$ is performed to reproduce the pure-gauge
Lattice data and $T_0=270~\rm{MeV}$ is adopted in our work. Finally
$\kappa=0.2$ is used which gives reasonable values for pressure for the
temperature range used here at zero baryon density as compared to full
Lattice QCD computations.

For simplicity, electrons are considered as free non-interacting
fermions \cite{buballa} and the corresponding thermodynamic potential
is,

\beq
\Omega_e= -(\frac{{\mu_e}^4}{12\pi^2} + \frac{{\mu_e}^2T^2}{6} +
            \frac{7\pi^2T^4}{180})
\eeq

\noindent
where, $\mu_e$ is the electron chemical potential.

\section{Results and Discussions} \label{sc.results}

The thermodynamic potential $\Omega$ is extremised with respect to
the scalar fields under the condition $\mu_d$=$\mu_u$+$\mu_e$ and 
$\mu_s$=$\mu_d$. The equations of motions for the mean fields 
$\sigma_u$, $\sigma_d$, $\sigma_s$, $\Phi$ and $\bar{\Phi}$ for any
given values of temperature $T$, quark chemical potential $\mu_q$
and electron chemical potential $\mu_e$ are determined through the
coupled equations,

\begin{equation}
\frac{\partial\Omega}{\partial\sigma_u}=0,\quad\frac{\partial\Omega}
{\partial\sigma_d}=0, 
\quad\frac{\partial\Omega}{\partial\sigma_s}=0
\quad\frac{\partial\Omega}{\partial\Phi}=0,\quad\frac{\partial\Omega}
{\partial\bar{\Phi}}=0.
\label{minmz}
\end{equation}

\begin{figure}[!htb]
\subfigure[]
{\includegraphics[scale=0.8]{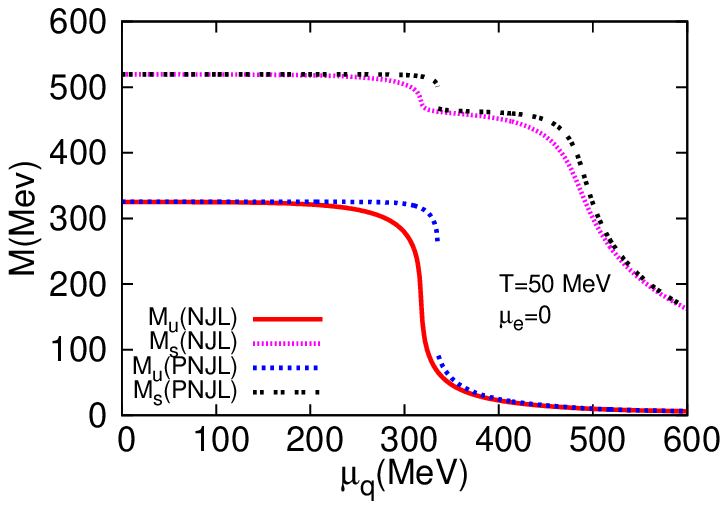}
\label{fg.mass_mue0}}
\subfigure[]
{\includegraphics[scale=0.8]{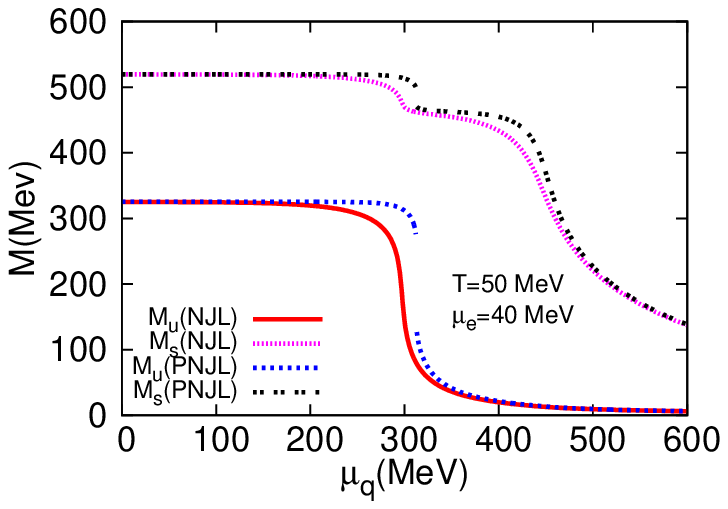}
\label{fg.mass_mue40}}
\caption{Constituent quark masses as functions of $\mu_q$
for (a) $\mu_e=0$ MeV and (b) $\mu_e=40$ MeV, at $T=50$ MeV.}
\label{fg.constmas}
\end{figure}

In Fig.\ref{fg.constmas}, we show the typical variation of constituent
quark masses as a function of $\mu_q$, for two representative values of
electron chemical potential $\mu_e=0$ MeV and $\mu_e=40$ MeV, with a
fixed temperature $T= 50$ MeV. At this temperature, both $m_u$ and $m_s$
in the PNJL model, show a discontinuous jump at around $\mu_q = 350$ 
MeV indicating a first order phase transition. The jump in $m_s$ is 
smaller,
and is actually a manifestation of chiral transition in the two flavor
sector, arising due to the coupling of the strange condensate to the
light flavor condensates. On the other hand in the NJL model the quark
masses show a smooth variation at this temperature, indicating a
crossover. It is important to note that the constituent mass of the
strange quark goes down to the current mass at a larger $\mu_q$ in both
the models, leading to sort of a second crossover at around
$\mu_q = 500$ MeV. This will have important implications for some of
the thermodynamic observables as we discuss below.

\begin{figure}[!hbt]
\subfigure[]
{\includegraphics[scale=0.9]{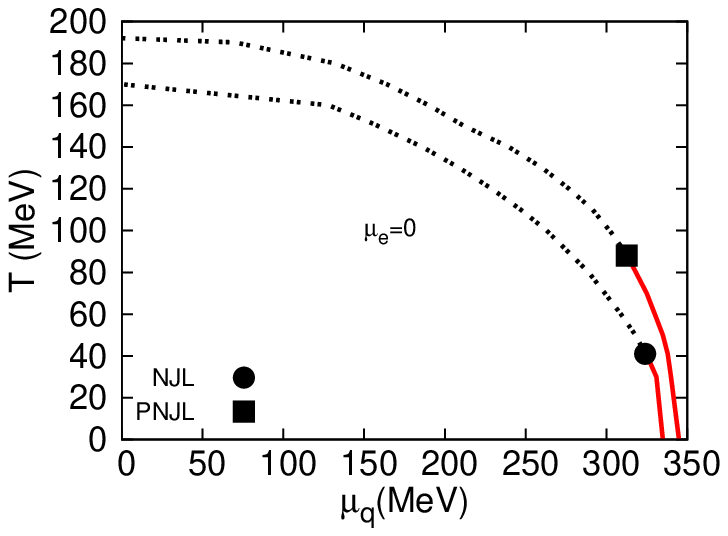}
\label{fg.njl_pnjl_ph_0}}
\subfigure[]
{\includegraphics[scale=0.9]{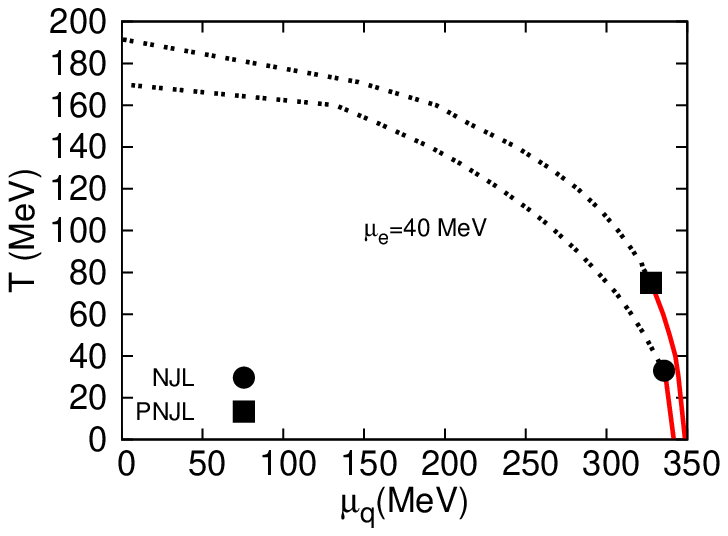}
\label{fg.njl_pnjl_ph_40}}
\caption{Comparison of phase diagram in NJL and PNJL model at 
$\beta-$equilibrium for (a) $\mu_e$=0 ; (b) $\mu_e$=40. The solid circle
and square represent the CEP for NJL and PNJL model respectively.}
\label{fg.njl_pnjl_ph}
\end{figure}

The phase diagrams for NJL and PNJL models are obtained from the 
behavior of the mean fields, and are shown in
Fig. \ref{fg.njl_pnjl_ph_0} and Fig. \ref{fg.njl_pnjl_ph_40} for 
$\mu_e=0$ MeV and $\mu_e=40$ MeV respectively. As is evident from the
figures, the broad features of the phase diagrams remain same in
all cases. The difference between the NJL and PNJL models arise
mainly due to the Polyakov loop, whose presence is primarily
responsible for raising the transition/crossover temperature in the
PNJL model. Thus the CEP for PNJL model occurs at slightly higher $T$
and lower $\mu_q$ compared to NJL model. Note that the phase diagram
with $\mu_e=0$ MeV is identical to the case  without
$\beta-$equilibrium \cite{paramita}. This is because the minimization
conditions (\ref{minmz}) are independent of the electrons except through
the $\beta-$equilibrium conditions. However this is true only so far as
the phase diagram is concerned. Various other physical quantities are
found to differ even for $\mu_e=0$ as discussed below. For non-zero
$\mu_e$ we find a slight lowering of the temperature for the CEP by
about $10~\rm{MeV}$. This is an important quantitative difference
between the physics of neutron stars and that of compressed baryonic
matter created in the laboratory. It is worth to mention that the
CEP we have obtained corresponds to chiral phase transition.
Generally in standard QCD phase diagram chiral and deconfinement
phase transition are shown by a single boundary. However 
it was argued and elucidated that the two transition lines in 
$T-\mu_B$ plane are distinct \cite{mclerran,fukushima_pd} .
In this context we would like  to mention that in \cite{shao} 
QCD phase diagram has been studied both for isospin asymmetric and symmetric
situations, although they have not considered the $\beta$- equilibrium 
scenario. The authors used a two equation 
of state model, non-linear Walecka model to describe 
hadronic sector and (P)NJL model for quark sector. It has been shown in
\cite{shao} that CEP remain unaffected by the isospin asymmetry and the
authors found it to be quite generic for a two EOS model.
\begin{figure}[!htb]
\subfigure[]
{\includegraphics[scale=1.0]{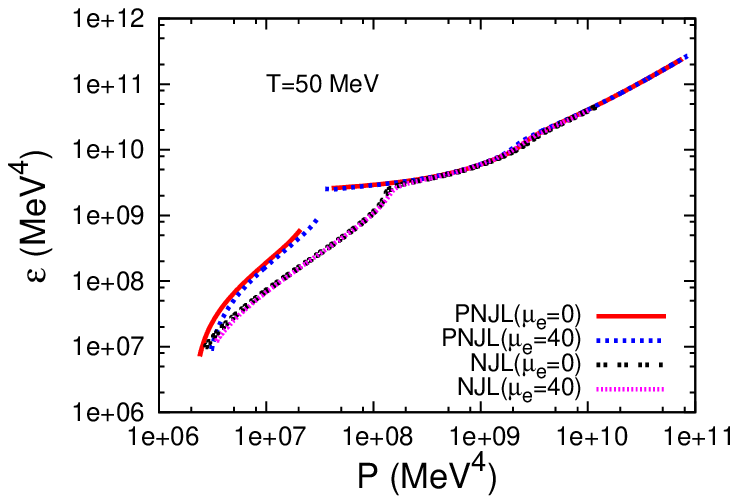}
\label{fg.EOS_compare}}
\subfigure[]
{\includegraphics[scale=1.0]{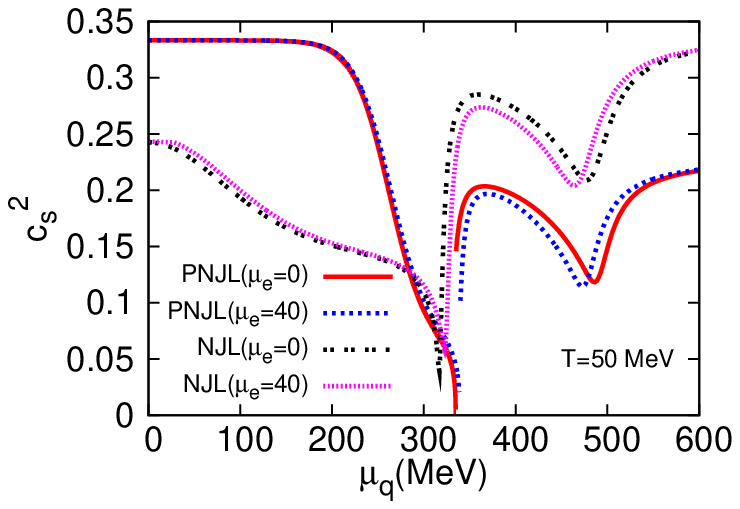}
\label{fg.csqr}}
\caption{ (a) Equation of state and (b) isentropic speed of sound, for
NJL and PNJL models at $T=50$ MeV.}
\label{fg.eos}
\end{figure}

The system under investigation can be characterized primarily by the 
behavior of the EOS. Generally for a many body system, increase in
pressure at large densities is indicative of a repulsive behavior of the
interaction at large densities (large $\mu_q$) or short distances and an
attractive nature at larger distances or lower densities
\cite{iwasaki,molodtsov}. Consequentially the energy density will show
similar behavior. The resulting EOS given by the variation of pressure
$P$ with energy density $\epsilon$, is shown in
Fig. \ref{fg.EOS_compare} at $T=50$ MeV, for both NJL and PNJL models,
for the two representative electron chemical potentials. Here again for
the PNJL model there exists a discontinuity due to a first order nature
of the transition, whereas for NJL model the EOS is smooth. Beyond this
region a smaller steepening in $\epsilon$ is visible, that occurs due to
second crossover feature noted above as the strange quark condensate 
starts to melt. A possible implication for this small surge may be that
in a strange quark star, at a given central density, the pressure would
be somewhat lesser than the situation without this surge.

Generally, the EOS can be used to study the dynamics of neutron star
and that of heavy-ion collisions through the respective flow equations.
The main differences would be due to the presence of $\beta-$equilibrium
and the back reaction of the non-trivial space-time metric on the EOS
for neutron stars. Such a comprehensive comparative study will be taken
up in a later work.

In Fig. \ref{fg.csqr}, the variation of the isentropic speed
of sound squared $c_s^2=\partial P/\partial \epsilon$ is plotted
against $\mu_q$ at $T=50$ MeV. In the NJL model the $c_s^2$ starts
from a non-zero value, steadily decreases and then shows a sharp
fall around the crossover region at $\mu_q \sim 320$ MeV. This is
followed by a sharp rise, a dip and then approaches the ideal gas
value of 1/3. In contrast the $c_s^2$ in the PNJL model starting from the
ideal gas value remains almost constant up to $\mu_q \sim 200$ MeV and
then falls sharply to almost zero. This is followed by a discontinuous
jump, a similar dip at $\mu_q \sim 500$ MeV and a gradual approach to
a non-zero value quite different from the ideal gas limit.

The difference at $\mu_q=0$ MeV occurs specifically due to the Polyakov
loop which suppresses any quark-like quasi-particles. As a result the
$c_s^2$ is completely determined by the ideal electron gas. On the
other hand those quasi-particles with heavy constituent masses tend to
lower the $c_s^2$ in the NJL model.
The difference at the transition region is again mainly due to the
discontinuous phase transition in PNJL model which leads to $c_s^2$
almost going down to zero, and a crossover in the NJL model where
$c_s^2$ is small but non-zero. In \cite{glend92} it was noted
that for two conserved charges, pressure is not constant any more in the
mixed phase, rather its variation becomes slower, resulting in a smaller
but non-zero speed of sound. In our computation though we do not find
$c_s^2$ exactly equal to zero, but to confirm such an effect we need a
full space-time simulation of the mixed phase through the process of
bubble nucleation which is beyond the scope of the present work.

 In both the models the dip around $\mu_q=500$ MeV arises due to the
behavior of the strange quark condensate as discussed earlier. If it
were possible to achieve such extremely high densities in heavy-ion
experiments, then such a dip would slow down the flow and would result 
in a larger fire ball life time. At even higher $\mu_q$ the
$c_s^2$ in NJL model approaches the free field limit quite fast but in
the PNJL model it still remains quite low due to the non-trivial
interaction brought in by the Polyakov loop. It would be interesting to
study the implication of slow speed of sound inside the core of a
neutron star.

\begin{figure}[!htb]
\subfigure[]
{\includegraphics[scale=0.6]{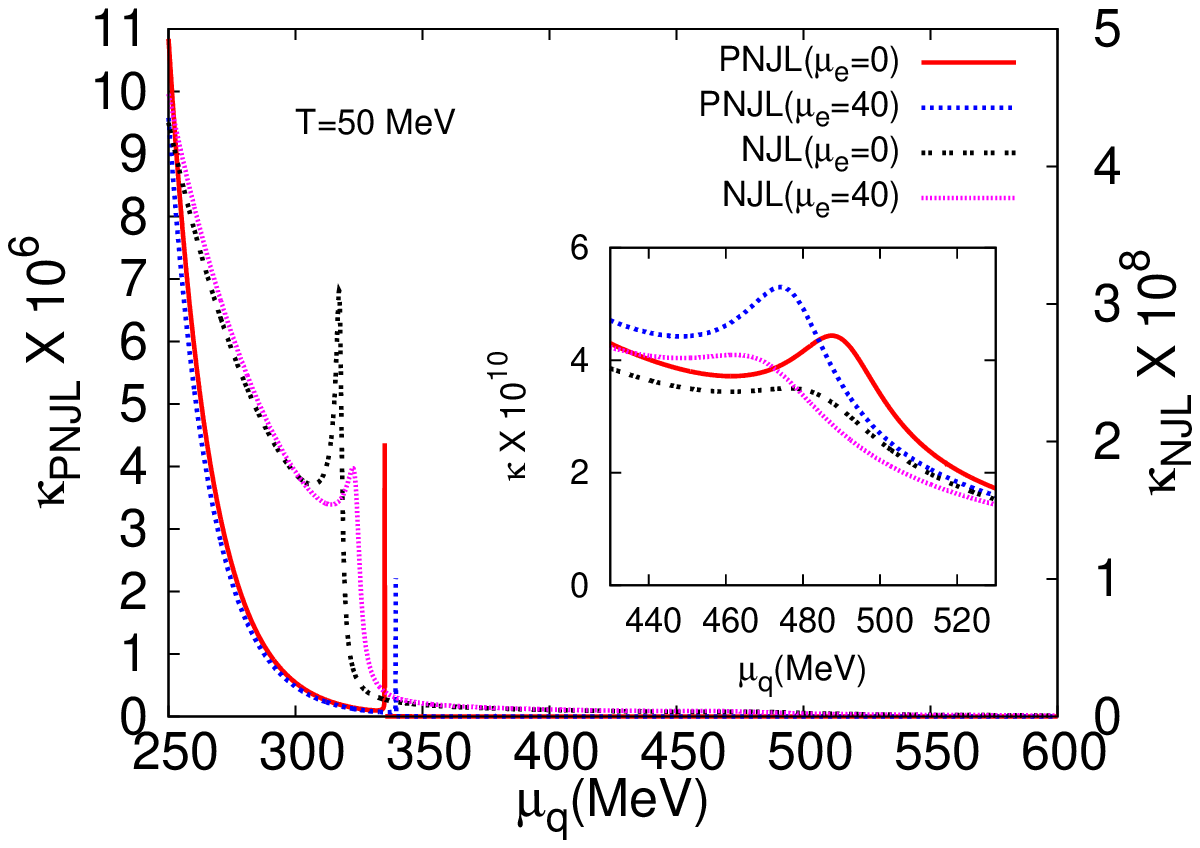}
\label{fg.kappa}}
\subfigure[]
{\includegraphics[scale=1.0]{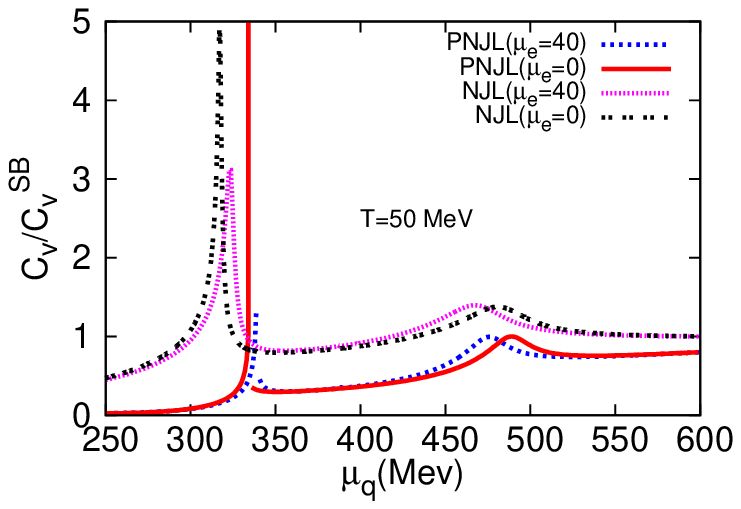}
\label{fg.sp_heat}}
\caption{ (a) Variation of compressibility $\kappa$ with $\mu_q$. The 
peak around $\mu_q=500$ MeV is shown in the inset where $\kappa$
represents the compressibility in both NJL and PNJL models.
(b) Variation of specific heat scaled by its Stefan Boltzmann value.}
\vskip 0.1 in
\end{figure}

Commensurate with the relative stiffening of the equation of state we
find that the compressibility $\kappa=\frac{1}{n_q^2}
{\left(\frac{\partial n_q}{\partial \mu_q}\right)}_T$, where $n_q$ is
the quark number density, behaves accordingly. While $\kappa$ in the NJL
model is found to be higher than that of the PNJL model in the hadronic
phase, it is just the opposite in the partonic phase as shown in
Fig. \ref{fg.kappa}. In the NS scenario this would mean that the core
of the star would be much softer compared to the crust if described by
the PNJL model rather than the NJL model. 

The variation of the specific heat
$C_V=T{\left(\frac{\partial s}{\partial T}\right)}_V$, where
$s=\left(\frac{\partial P}{\partial T}\right)$ is the entropy density of
the system, is shown in Fig. \ref{fg.sp_heat}. For a crossover (here in
NJL model) the specific heat shows a peak. For a first order transition
(here in the PNJL model) the $C_V$ is discontinuous. Also we see that
the specific heat in the PNJL model is lower than that in the NJL model
for a general variation of $\mu_q$ and $\mu_e$. A system described by
the PNJL model is thus less susceptible to changing temperature than
that described by the NJL model.

The variation of compressibility and specific heat shown here also
captures the signature of a phase transition in the PNJL model and a
crossover in the NJL model. Both compressibility as well as
specific heat are second derivatives of $\Omega$ and represent
respectively the quark number fluctuations and energy fluctuations
\cite{iwasaki}. Discontinuity in compressibility as well as
specific heat indicates a first order phase transition for the PNJL
model. At $\mu_q \sim 500$ MeV, both the models exhibit a small peak due
to the onset of melting of the strange quark condensate.

\begin{figure}[!htb]
\subfigure[]
{\includegraphics[scale=1.0]{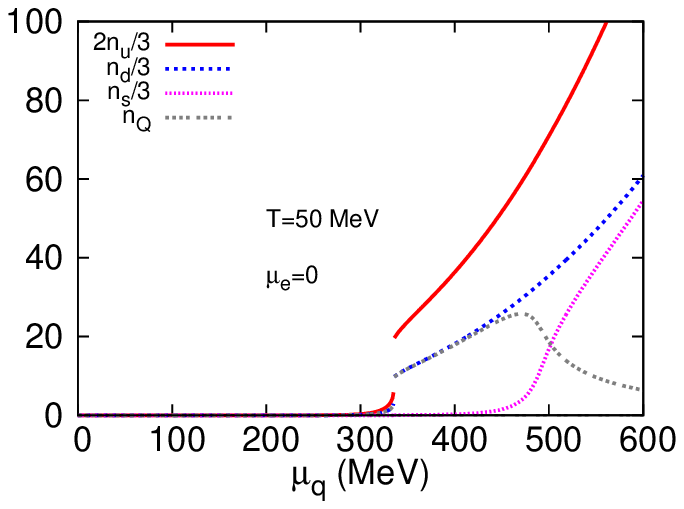}
\label{fg.nQ_mue0}}
\subfigure[]
{\includegraphics[scale=1.0]{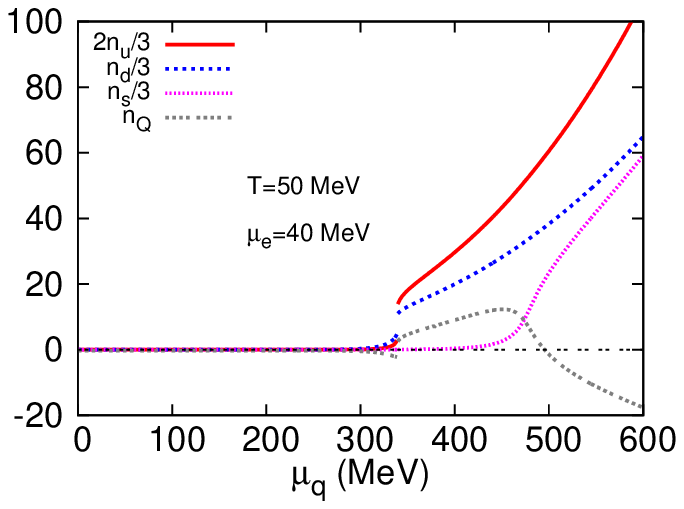}
\label{fg.nQ_mue40}}
\caption{Total charge and quark number densities scaled by $T^3$ as
a function of quark chemical potential in the PNJL model}
\label{fg.netcharge}
\end{figure}

We now consider the net charge density given by
$n_Q=\frac{2}{3}{n_u}-\frac{1}{3}{n_d}-\frac{1}{3}{n_s}-{n_e}$, where
the number density of individual quarks and electrons are obtained
from the relations,
$
n_u=\frac{\partial\Omega}{\partial\mu_u},~
n_d=\frac{\partial\Omega}{\partial\mu_d},~
n_s=\frac{\partial\Omega}{\partial\mu_s},~{\rm and}~
n_e=\frac{\partial\Omega_e}{\partial\mu_e}
$.
For $\mu_e=0$, $n_e=0$ and we have $n_u$ = $n_d$. At large $\mu_q$,
the number density $n_s$ of strange quarks become almost equal to the
light quark number densities as the constituent masses of strange quarks
are reduced significantly. So the net charge density $n_Q$ will be close
to zero and the system will become charge neutral asymptotically as
shown in Fig. \ref{fg.nQ_mue0}. At small $\mu_q$, $n_Q << 1$ as the
individual number densities themselves are exceedingly small. In fact
this feature continues till the transition region where the light
constituent quark masses drop sharply giving rise to non-zero number
densities. Therefore $n_Q$ shows a non-monotonic behavior, rising from
almost zero it reaches a maxima at certain $\mu_q$ determined mainly by
the melting of the strange quark condensate and thereafter decreases
steadily towards zero.

For higher $\mu_e$, charge neutral configuration is possible even at
non-zero moderate values of $\mu_q$. For small $\mu_q$, it is the $n_e$
which dominates and keeps $n_Q$ negative. As soon as $n_u$ becomes large
with increasing $\mu_q$, $n_Q$ goes through zero and becomes positive.
Now since $\mu_s$ and $\mu_d$ are greater than $\mu_u$ due to
$\beta-$equilibrium, both $n_s$ and $n_d$ start to grow faster with the
increase of $\mu_q$. Finally at some $\mu_q$ the net charge becomes zero
due to the mutual cancellation of $n_u$, $n_d$, and $n_s$, and
thereafter it remains negative for higher $\mu_q$ as $d$ and $s$ quarks
overwhelms the positively charged $u$ quark. The electron number
density is fixed for a fixed value of $\mu_e$ and $T$, and it is
negligible compared to the quark number densities at high $\mu_q$. The
behavior of $n_Q$ is similar for both PNJL and NJL model though the
actual values of the various chemical potentials for the charge
neutrality conditions vary. 

\begin{figure}[!htb]
\subfigure[]
{\includegraphics[scale=0.9]{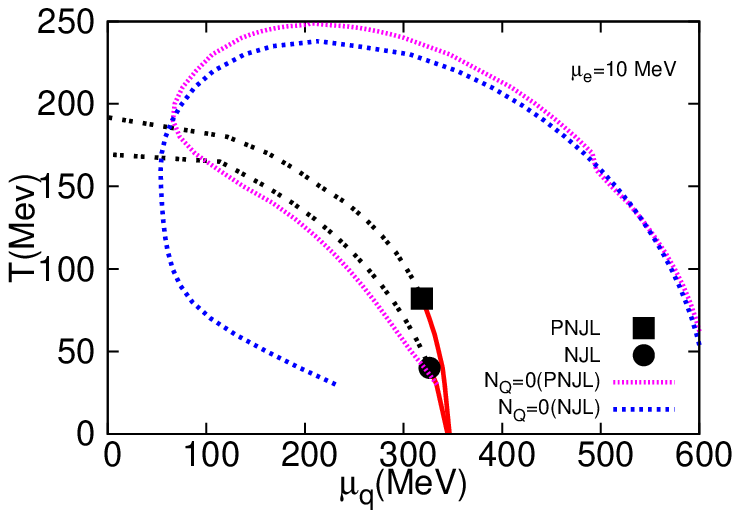}
\label{fg.njl_pnjl_10}}
\subfigure[]
{\includegraphics[scale=0.9]{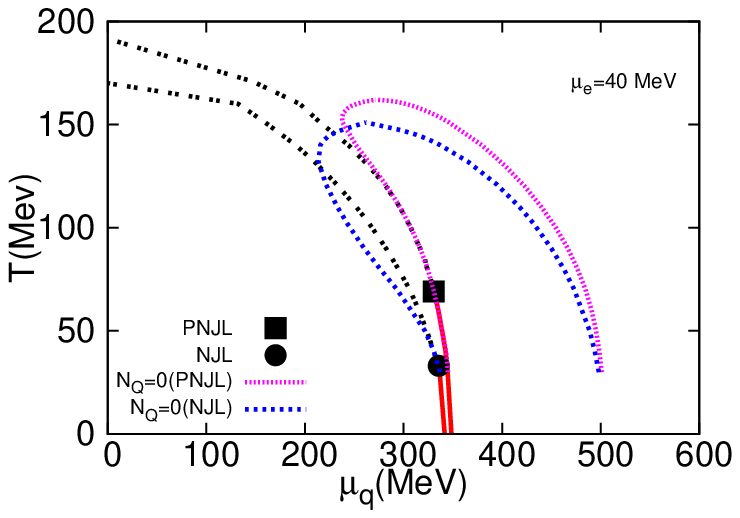}
\label{fg.njl_pnjl_40}}
\caption{Comparison of charge neutral trajectory in NJL and PNJL model
at (a) $\mu_e$=10 ; (b) $\mu_e$=40}
\label{fg.njl_pnjl_0chrg}
\end{figure}

Given that one may be interested in the charge neutral condition
{\it e.g.} in the case of neutron stars, in Fig. \ref{fg.njl_pnjl_0chrg}
the charge neutral trajectories for NJL model are compared with those of
PNJL model along with the phase diagrams. The trajectories are quite
interesting in that they are closed ones pinned on to the $\mu_q$ axis.
They start off close to $\mu_q = M_{vac}$, the constituent quark mass
in the model in vacuum. They make an excursion in the $T-\mu_q$ plane
and join back at a higher $\mu_q$. There is a maximum temperature $T_Q$
up to which the trajectory goes. Beyond this temperature no charge
neutrality is possible. Below this temperature we have essentially two
values of $\mu_q$ where charge neutrality occurs. There are significant
differences between the contours of NJL and PNJL model in the hadronic
phase. However beyond the transition and inside the deconfined region,
the differences subside as the Polyakov loop relaxes the confining
effect leading to the PNJL model behaving in a similar way to that of
the NJL model. 

 The behavior of the charge neutral contour is highly dependent on
$\mu_e$. With increasing $\mu_e$ the contour gradually closes in towards
the transition line. For a given $T$ there are two $\mu_q$ values where
charge neutrality is obtained $-$ one on the hadronic side and one on
the partonic side. As a result of the closing in of the contour, these
two values come closer to the transition line from opposite sides with
increasing $\mu_e$. Higher the $\mu_e$ closer we are to the transition
region. Now suppose we are looking for an isothermal evolution of a
system, or the isothermal configuration of a system such as the NS.
Given the constraint of charge neutrality we would have a varying
$\mu_e$ as the density profile changes. Similarly if $\mu_e$
is held constant then charge neutrality would not allow the temperature
to remain fixed throughout and the evolution would take place along the
contours described above. So in general a combination of $T$ and
$\mu_e$ is expected to maintain charge neutrality in a given system.
A practical picture of NS which has a profile of low density crust to
gradual increase in density to have a highly condensed core would be
that there is a complex profile for temperature and $\mu_e$ inside the
NS. In fact if there exist a hadron-parton boundary, it may be either
with high temperature or high electron density.

\begin{figure}[!htb]
\subfigure[]
{\includegraphics[scale=0.9]{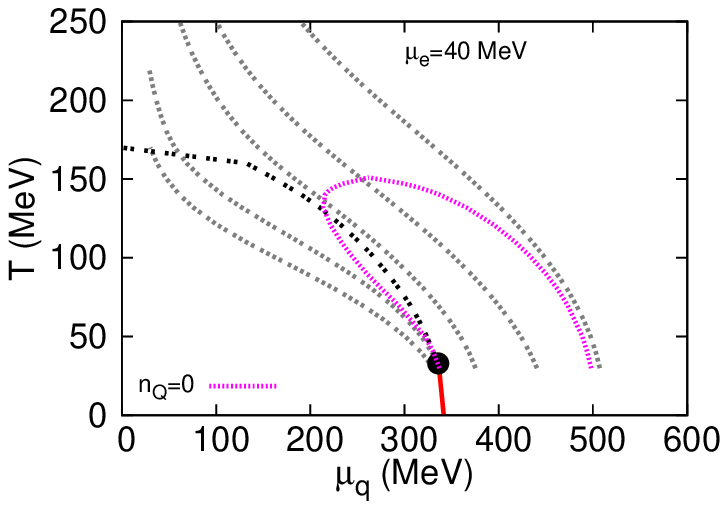}
\label{fg.rhocontour_0}}
\subfigure[]
{\includegraphics[scale=0.9]{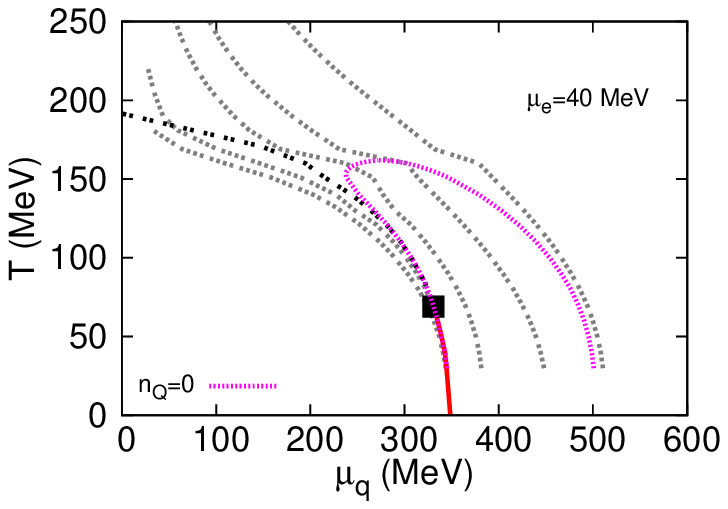}
\label{fg.rhocontour_40}}
\caption{The contour of scaled baryon number density $n_B/n_0$;
(scaled by normal nuclear matter density) along with phase diagram at 
$\mu_e$=40 for (a) NJL model and (b) for PNJL model;
(From left $n_B/n_0 =$ 0.5, 1, 3, 5, 10 respectively)}
\label{fg.rhocontour}
\end{figure}

To contemplate this scenario in the light of the baryon densities
achieved we plot the contours for constant baryon densities, scaled by
the normal nuclear matter density ($n_0 = 0.15fm^{-3}$) in
Fig. \ref{fg.rhocontour} for $\mu_e=$40 MeV. The charge neutral
trajectories are also plotted along with the phase boundary.
Obviously with increasing baryon (quark) chemical potential baryon
density would increase. What is interesting is the fact that high
densities can also occur for lower chemical potential if the temperature
is higher. For both NJL and PNJL model at and above 3 times nuclear
matter density the matter seems to be always in the partonic phase. A
little below this density matter may be in partonic phase if it is at
high temperature otherwise in the hadronic phase at low temperature.
Thus the actual trajectory on the phase diagram would determine whether
a hadron-parton boundary in the NS is in the mixed phase or in a state
of crossover. Within the range of the charge neutral contour we find the
baryon density increasing from a very small value to almost 10 times the
normal nuclear matter density. If $\mu_e$ is increased further the
baryon densities would also be much higher for a given $\mu_q$. So
if we assume local charge neutrality as well as isothermal profile
along a hadron-parton phase boundary, the baryon density close to the
phase boundary may be too large. On the other hand for reasonable 
densities close to the phase boundary it would be impossible to
maintain local charge neutrality along an isothermal curve. In that
case it may be possible that the charge neutrality condition takes the
system around the CEP to hold on to a reasonable density in the phase
boundary region. This leads us to speculate that the transition in a NS
itself may also be a cross-over, quite unlike the picture in most of the
studies of NS.

\begin{figure}[!htb]
\subfigure[]
{\includegraphics[scale=0.9]{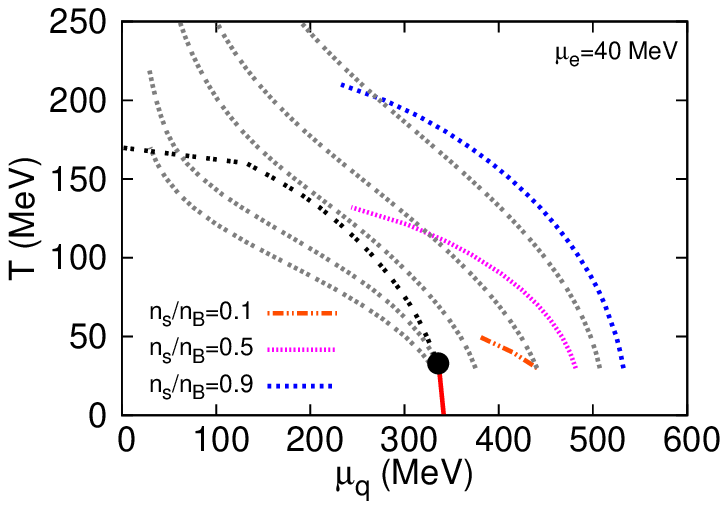}
\label{fg.strange_40_njl}}
\subfigure[]
{\includegraphics[scale=0.9]{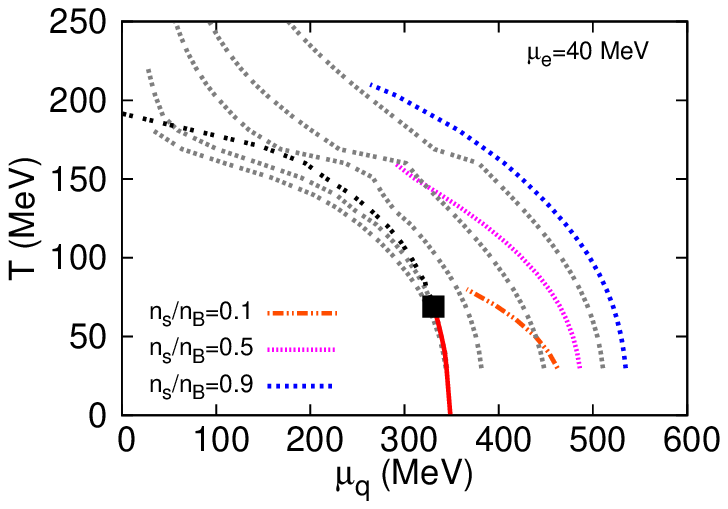}
\label{fg.strange_40_pnjl}}
\caption{The contour of net strangeness fraction ($n_s/n_B$) along with
$n_B/n_0$ at $\mu_e$=40 for (a) NJL model and
(b) PNJL model; the values of $n_B/n_0$ are 0.5, 1, 3, 5 and 10
(from left). }
\label{fg.netstrng}
\end{figure}

The net strangeness fraction ($n_s/n_B$) along with $n_B/n_0$ is shown
in Fig. \ref{fg.netstrng}. For a given temperature, there is a critical
$\mu_q$ below which there is no net strangeness formation. At the
critical $\mu_q$ a non-zero $n_s/n_B$ occurs depending on the $T$. This
strangeness fraction continues to appear at lower $T$ for some higher
$\mu_q$. So a given strangeness fraction can occur only upto a certain
critical temperature. The intersection of lines of constant baryon
density and strangeness fraction indicates the possibility of evolution
of a system to higher (lower) strangeness fraction with increase
(decrease) of $T$ at a constant density. In the range of 5 - 10 times
nuclear matter density we see that the strangeness fraction is
increasing significantly towards unity indicating a possibility of
formation of quark matter with almost equal number of u, d and s quarks.
Similar results have also been found in other model studies
\cite{ghosh_star1}. Chunks of matter with $n_s/n_B=1$, called
strangelets is expected to be stable (metastable up to weak decay)
relative to nuclear matter in vacuum \cite{farhi}. Investigation of
these and various other properties of strange matter would be undertaken
in future.

\begin{figure}[!htb]
\subfigure[]
{\includegraphics[scale=0.9]{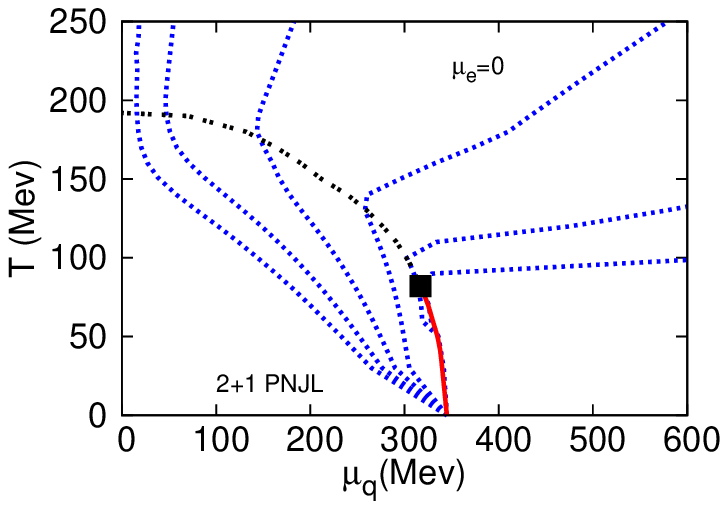}
\label{fg.contour_0}}
\subfigure[]
{\includegraphics[scale=0.9]{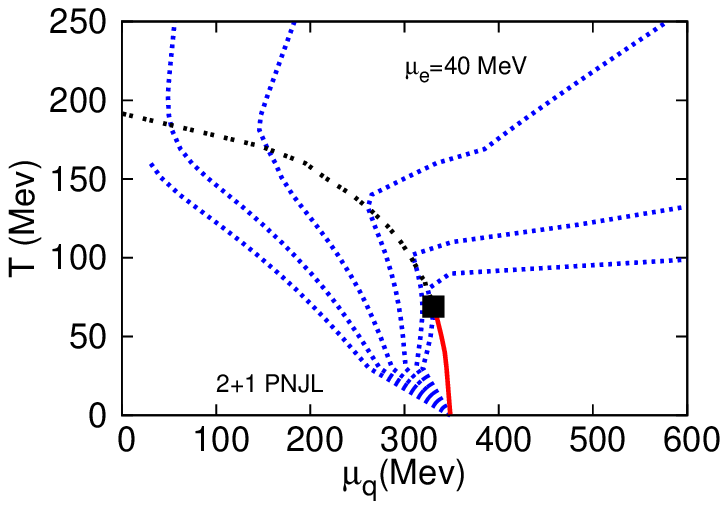}
\label{fg.contour_40}}
\subfigure[]
{\includegraphics[scale=0.9]{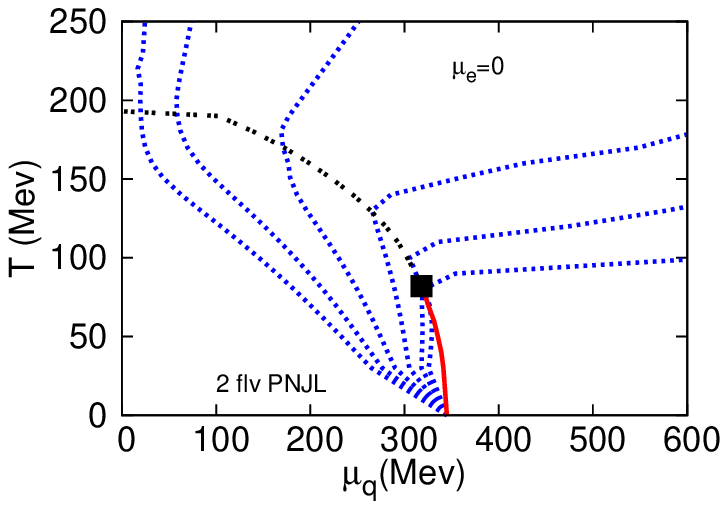}
\label{fg.contour_s0}}
\subfigure[]
{\includegraphics[scale=0.9]{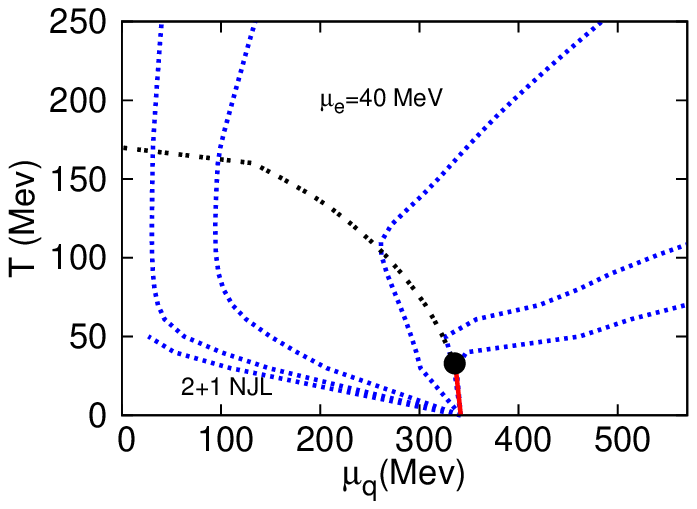}
\label{fg.contour_40_njl}}
\caption{The isentropic trajectories along with phase diagram for
(a) at $\mu_e$=0, 2+1-flavor PNJL,
(b) at $\mu_e$=40, 2+1-flavor PNJL,
(c) at $\mu_e$=0, 2-flavor PNJL and
(d) at $\mu_e$=40, 2+1-flavor NJL model.
$s/{n_B}$=300,100,30,10,5,3.5 (from left).}
\label{fg.adiabat}
\end{figure}

Usually the hydrodynamic evolution of a system is expected to follow
certain adiabat along which the entropy per baryon number ($s/n_B$) is
a constant quantity. Among the various adiabats the system would choose
one given its initial conditions. In the context of NS, a fixed entropy
per baryon is expected in a proto-neutron star as well which is very
different from a cold neutron star. It is usually hot and rich in
leptons {\it i.e.} electrons and trapped neutrinos. Few seconds after
birth, the matter in the core of a hot NS has almost constant lepton
fraction (0.3 -0.4) and entropy per baryon (1 - 2, in units of Boltzmann
constant) \cite{burrows,gondek}. The question as to whether the later
evolution of the NS can be described to be one close to an adiabat
is a matter of debate. On the other hand the commonly used approach of
an isothermal evolution looks not quite favorable according to the
above discussion on charge neutrality condition.

 The behavior of $s/n_B$ in a plasma and in a hadron gas was analyzed
within the framework of an extended Bag model by \cite{leonidov-prd50}.
A case study of such adiabats was done in NJL model in
\cite{scaveniusprc64}. It was found that unlike the prescription of
adiabats meeting at the CEP given by \cite{shuryakprl81}, they meet
close to the critical value of $\mu_q$ at $T=0$ which is incidentally
equal to the constituent quark mass $M_{vac}$ in the model in vacuum.
It was argued in \cite{scaveniusprc64} that as $T \rightarrow  0$,
$s \rightarrow 0$ by the third law of thermodynamics. Hence in order to
keep $s/n_B$ constant, $n_B$ should go to zero. This condition is
satisfied when $\mu_q=M_{vac}$ of the theory. These authors also found
similar results for the linear sigma model. In the PNJL model the
introduction of Polyakov loop produced a slight change in the
configuration of the adiabats \cite{kahara-prd}. The constraint on
the strangeness number to be zero also was found not to have a very
significant effect \cite{fukuisen}.

The corresponding picture of isentropic trajectories with the condition
of $\beta-$equilibrium is shown in Fig. \ref{fg.adiabat}. Four cases
are depicted here. Fig. \ref{fg.contour_0} and Fig. \ref{fg.contour_40}
show the cases with 2+1 PNJL model at $\mu_e=0$ MeV and $\mu_e=40$ MeV
respectively. From these two figures we find that the electron density
does not have a significant effect on the isentropic trajectories. This
means that the quark degrees of freedom seem to have dominant effect
in entropy over the electrons. The case with $n_s=0$, {\it i.e.}
effectively for a $2-$flavor system is shown in Fig. \ref{fg.contour_s0}.
In general the situation is similar. For small $\mu_q$ there is almost
no change in Fig. \ref{fg.contour_s0} and Fig. \ref{fg.contour_0} as 
both the cases are identical to 2 flavors. At intermediate values
strange quarks start to pop out. Now the contours in
Fig. \ref{fg.contour_s0} appear to be shifted and bent towards higher
$\mu_q$. This is because for 2 flavors, a given baryon number density
appears at a higher $\mu_q$ than that for 2+1 flavors. Hence to get a
fixed $s/n_B$ the $\mu_q$ required is also higher. At even higher
$\mu_q$ the thermal effects are negligible and hence $s/n_B$ become
almost independent of the degrees of freedom. Thus again the contours
become identical.

The results in the NJL model are significantly different from that
of the PNJL model as can be seen by comparing the PNJL results with that
of the NJL model shown in Fig. \ref{fg.contour_40_njl}. Even for low $T$
and $\mu_q$ there is a significant entropy generation as there is no
Polyakov loop to subdue the same. Similar differences continue to appear
even in the partonic phase.

Considering a system that has been compressed to a few times the nuclear
matter density it can try to relax back to lower densities along the 
adiabats. Interestingly the isentropic trajectories in the high density
domain seem to behave as isothermals in the PNJL model. However as soon
as the system converts into the hadronic phase, the adiabats drive it
to a steep fall in temperature. We would like to mention that for a
hadronic proto neutron star with beta-equilibrated nuclear matter with
nucleons and leptons in the stellar core, the EOS evaluated in
Bruckner$-$Bethe$-$Goldstone theory, was found to be similar for both
isothermal and isentropic profiles \cite{burgio2}.
\par
In Ref.\cite{costa_parameter} isentropic trajectories were obtained in 
PNJL model without the constraint of $\beta$- equilibrium, for two
different sets of parameters corresponding to
an ultraviolet cutoff in the zero temperature integrals only (case I)
and the same in all integrals (Case II). Here we considered only
the first case for regularisation and find similar results.

While the possibility that a neutron star can be described using
adiabatic conditions is a point to be pondered about, we note here that
an excursion of the phase diagram of a $\beta-$equilibrated matter is
highly possible even in heavy-ion collisions to some extent. This is
because both the isentropic lines as well as the characteristics of the
phase boundary are quite similar for a wide variation of $\mu_e$ and
$\mu_q$. At the same time one should remember that in the laboratory
conditions $n_s$ is strictly zero. Anyway if a system is found to have
travelled along an adiabat with $s/n_B \simeq$ 3 to 4, it has most
probably traversed close to the CEP. One can therefore try to correlate
different observables like the enhancement of fluctuations of conserved
charges and $s/n_B$ to be in the above range to study the approach
towards the CEP in heavy-ion collisions.
 
\section{Summary and Conclusion} \label{sc.conclusion}

In this paper we have studied the 2+1 flavor strongly interacting matter
under the condition of $\beta-$equilibrium. We have presented a
comparative study of NJL versus PNJL model. The phase diagrams in these
two models are broadly similar, but quantitatively somewhat different.
The presence of the Polyakov loop delays the transition for larger
values of temperature for a given quark chemical potential. As a result
the CEP in the PNJL model is almost twice as hot as that in the NJL
model. We have illustrated characteristics of the phase diagram with
the behavior of some thermodynamic quantities like the constituent
mass, compressibility, specific heat, speed of sound and the
equation of state for $\mu_e=0$ MeV and $\mu_e=40$ MeV at $T$=50 MeV.
We found striking differences between the NJL and PNJL model in terms
of the softness of the equation of state in the hadronic and partonic
phases.

The behavior of electric charge and baryon densities in the two models
also differ in the hadronic phases to some extent. The differences
become less with increasing electron density. We explained how the
charge neutral trajectory is important in deciding the path along which
the core of NS can change from hadronic to quark phase. For all values
of $\mu_e$ we find that the contours are all closed ones and give a
restricted range of temperature and densities that are allowed. We
speculated a possible scenario in which the quark-hadron transition in
a NS would be a crossover. Again the baryon density contours seemed to
suggest that if a system has baryon density three times the nuclear
matter density it is quite surely in the partonic phase. We also
found that the strangeness fraction increases steadily with increasing
baryon density implying a possibility of having a strange NS.

The isentropic trajectories were obtained along which a system
in hydrodynamic equilibrium is expected to evolve. The adiabats flow
down from high temperature and low density towards low temperature and
$\mu_q=M_{vac}$, the constituent quark mass in vacuum. The adiabats then
steeply rise along the transition line, thereafter goes towards higher
densities with almost a constant slope. For small $s/n_B$ ratio the
slope is so small that the isentropic trajectories almost become 
isothermal trajectories as well. 

To summarize the scenario inside neutron stars we note that inside a
newly born NS the temperature drops very quickly and gives rise to a
system of low temperature nucleonic matter which may also be populated
by hyperons and strange baryons due to high density near the core. The
star is assumed to be $\beta-$equilibrated and charge neutral. Now it
is possible that due to some reason, {\it e.g.} sudden spin down, this
nucleonic matter will start getting converted to predominantly two
flavor quark matter within strong interaction time scale. This
transition would start at the center and a conversion front moving
outward will convert much of the central region of the star. Along the
path of the conversion front, each point inside the star may lie on an
isentropic trajectory. Gradually this system of predominantly 2 flavor
quark matter will get converted to strange quark matter through
weak interactions and finally a $\beta-$equilibrated charge neutral
strange quark matter will be produced. The strangeness production occurs
mainly through non-leptonic decay \cite{ghosh-npa}, the system is
expected to lie on a constant density line and move towards the point
with highest strangeness possible at that density. Finally the
semi-leptonic processes will take over and system will then evolve along
a $\beta-$equilibrated charge neutral contour.  

The natural extension of the work is to obtain the detailed evolution
of a family of neutron stars starting with different initial conditions
and gravity effects incorporated. We hope to report the study in a future
publication. It would also be important to consider colored exotic 
states like diquarks \cite{bentz-npa} that may arise at high densities.

\section{Acknowledgement}
S.M would like to thank CSIR for financial support. A.B. thanks UGC
(DRS \& UPE) and DST for support. R.R. thanks DST for support. We would
like to thank Anirban Lahiri, Paramita Deb and Sibaji Raha for useful
discussion and comments.



\end{document}